\documentclass[a4paper,12pt, epsfig]{article}
\usepackage{epsfig}
\usepackage{amssymb}
\usepackage{amsfonts}
\usepackage{amsmath}
\usepackage{graphicx,subfigure}

\newskip\humongous \humongous=0pt plus 1000pt minus 1000pt

\newif\ifdtup

\catcode`\@=11

\@addtoreset{equation}{section}
\def\theequation{\thesection.\arabic{equation}}

\def\@normalsize{\@setsize\normalsize{15pt}\xiipt\@xiipt
\abovedisplayskip 14pt plus3pt minus3pt%
\belowdisplayskip \abovedisplayskip
\abovedisplayshortskip \z@ plus3pt%
\belowdisplayshortskip 7pt plus3.5pt minus0pt}

\def\small{\@setsize\small{13.6pt}\xipt\@xipt
\abovedisplayskip 13pt plus3pt minus3pt%
\belowdisplayskip \abovedisplayskip
\abovedisplayshortskip \z@ plus3pt%
\belowdisplayshortskip 7pt plus3.5pt minus0pt
\def\@listi{\parsep 4.5pt plus 2pt minus 1pt
      \itemsep \parsep
      \topsep 9pt plus 3pt minus 3pt}}

\relax

\catcode`@=12

\catcode`\@=11

\def\section{\@startsection{section}{1}{\z@}{3.5ex plus 1ex minus
    .2ex}{2.3ex plus .2ex}{\large\bf}}

\def\thesection{\arabic{section}}
\def\thesubsection{\arabic{section}.\arabic{subsection}}

\def\appendix{\setcounter{section}{0}
  \def\thesection{Appendix \Alph{section}}
  \def\thesubsection{\Alph{section}.\arabic{subsection}}
  \def\theequation{\Alph{section}.\arabic{equation}}}

\def\SymBoxes#1#2#3#4{\newdimen\un@t \un@t#3%
\raisebox{#1}{\rule{#2\un@t}{#4}\hskip-#2\un@t
\@tempdimb\un@t \advance\@tempdimb by-#4\@tempcntb#2\relax%
\@whilenum{\@tempcntb>0}\do{
\rule{#4}{\un@t}\hskip\@tempdimb \advance\@tempcntb by\m@ne}%
\hskip-#2\un@t \rule[\un@t]{#2\un@t}{#4}%
\rule[\un@t]{#4}{#4}\hskip-#4
\rule{#4}{\un@t}}\hskip-#4}                

\begin{document}

\newcommand{\beq}{\begin{equation}}
\newcommand{\eeq}{\end{equation}}
\newcommand{\bea}{\begin{eqnarray}}
\newcommand{\eea}{\end{eqnarray}}
\newcommand{\beas}{\begin{eqnarray*}}
\newcommand{\eeas}{\end{eqnarray*}}
\newcommand{\defi}{\stackrel{\rm def}{=}}
\newcommand{\non}{\nonumber}
\newcommand{\bquo}{\begin{quote}}
\newcommand{\enqu}{\end{quote}}
\def\de{\partial}
\def\Om{\ensuremath{\Omega}}
\def\Tr{ \hbox{\rm Tr}}
\def\H{ \hbox{\rm H}}
\def\HE{ \hbox{$\rm H^{even}$}}
\def\HO{ \hbox{$\rm H^{odd}$}}
\def\HEO{ \hbox{$\rm H^{even/odd}$}}
\def\HOE{ \hbox{$\rm H^{odd/even}$}}
\def\HHEO{ \hbox{$\rm H_H^{even/odd}$}}
\def\HHOE{ \hbox{$\rm H_H^{odd/even}$}}
\def\K{ \hbox{\rm K}}
\def\Im{ \hbox{\rm Im}}
\def\Ker{ \hbox{\rm Ker}}
\def\const{\hbox {\rm const.}}
\def\o{\over}
\def\im{\hbox{\rm Im}}
\def\re{\hbox{\rm Re}}
\def\bra{\langle}\def\ket{\rangle}
\def\Arg{\hbox {\rm Arg}}
\def\Re{\hbox {\rm Re}}
\def\Im{\hbox {\rm Im}}
\def\exo{\hbox {\rm exp}}
\def\diag{\hbox{\rm diag}}
\def\longvert{{\rule[-2mm]{0.1mm}{7mm}}\,}
\def\a{\alpha}
\def\dag{{}^{\dagger}}
\def\tq{{\widetilde q}}
\def\p{{}^{\prime}}
\def\W{W}
\def\N{{\cal N}}
\def\hsp{,\hspace{.7cm}}
\newcommand{\C}{\ensuremath{\mathbb C}}
\newcommand{\Z}{\ensuremath{\mathbb Z}}
\newcommand{\R}{\ensuremath{\mathbb R}}
\newcommand{\rp}{\ensuremath{\mathbb {RP}}}
\newcommand{\cp}{\ensuremath{\mathbb {CP}}}
\newcommand{\vac}{\ensuremath{|0\rangle}}
\newcommand{\vact}{\ensuremath{|00\rangle}                    }
\newcommand{\oc}{\ensuremath{\overline{c}}}
\def\theequation{\arabic{equation}}

\def\IR{{\mathbb R}}

\def\ka{\kappa}
\def\bc{{\bar c}}

\begin{titlepage}
\begin{flushright}
SISSA/26/2008/EP\\
ULB-TH/08-12\\
\end{flushright}
\bigskip
\def\thefootnote{\fnsymbol{footnote}}

\begin{center}
{\large {\bf
Metastable Black Saturns
  } }
\end{center}

\bigskip
\begin{center}
{\large  Jarah EVSLIN$^{1}$\footnote{\texttt{evslin@sissa.it}} and 
Chethan KRISHNAN$^{2}$\footnote{\texttt{chethan.krishnan@ulb.ac.be}}}
\end{center}

\renewcommand{\thefootnote}{\arabic{footnote}}

\begin{center}
\vspace{1em}
{\em  $^1${ SISSA,\\
Via Beirut 2-4,\\
I-34014, Trieste, Italy\\
\vskip .4cm
$^2$ International Solvay Institutes\\
Physique Th\'eorique et Math\'ematique\\
ULB C.P. 231, Universit\'e Libre de Bruxelles\\
B-1050, Bruxelles, Belgium  }
}
\end{center}

\noindent
\begin{center} {\bf Abstract} \end{center}
\noindent
Black Saturns have multiple horizons and so offer a testing ground for 
the ideas of black hole thermodynamics. In this note, we numerically scan for phases that are in equilibrium by extremizing total entropy in the 2-dimensional moduli 
space of stationary, singly rotating black Saturns with fixed total mass 
and angular momentum. On top of the known $T_H=T_R, \ \Omega_H=\Omega_R$ configurations, we find phases that do not balance the temperature and 
angular velocity of the ring and the hole. But these (and most of the 
balanced Saturns) go away when we demand that the system is metastable, by imposing that the Hessian of the entropy is negative definite. Metastablity occurs when the dimensionless total angular momentum lies in a narrow window $0.92457<j<0.92463$ of the 
thin ring branch. This is consistent with the expected range of 
classical stability of black Saturns and therefore may imply that thermal stability is tied to classical stability, in analogy with Gubser-Mitra in the translationally-invariant case.  We also comment on the possibility of constructing plasma configurations that are dual to black Saturns in AdS.

\vfill

\begin{flushleft}
\today
\end{flushleft}
\end{titlepage}

\hfill{}


\setcounter{footnote}{0}
5-dimensional gravity solutions 
\cite{ring,rings,Obers,ringstab,simp1,simp2,simp3,tomi,Figueras, Obers2, Rogatko, Yaz1, Yaz2, Yaz3} are 
fashionable for a number 
of reasons.  First of all, holography relates a 4-dimensional 
gauge theory to 5-dimensional gravity in an asymptotically AdS space and 
in particular maps 4-dimensional plasmas (Scherk-Schwarz reduced to an 
an annulus) to 5-dimensional 
black rings \cite{shiraz,shiraz2}.  Secondly, 5-dimensional gravity in 
asymptotically flat space is much richer than the 4-dimensional gravity 
of our universe.  In four
dimensions, a simple argument based on the positivity of energy and genus counting 
of two dimensional surfaces shows that the only admissible horizons have the topology of 
$S^2$. But in 5-D, the same positivity argument is far less restrictive. 
We can (and do) have not only black holes but also 
black rings and their composites, black Saturns \cite{Saturn} as 
solutions.  
Black Saturns, along with di-rings \cite{di,Iguchi:2007is, di2} and 
bi-rings \cite{bi,bi2}, 
are particularly interesting because superimposing two objects yields 
tunable parameters. Such ``hairy" black objects, which are not allowed in 
four dimensions, can be 
useful for understanding 
general features of quantum gravities with holographic descriptions.  For 
example, one can create abysses where quantum gravity effects become 
relevant at macroscopic scales \cite{bena} suggesting that perhaps 
general relativity breaks down not at the Planck scale as measured by a 
local observer, but rather as measured by an observer in the reference 
frame of the holographic screen. Another popular motivation for the study of higher dimensional black holes is that the Universe we live in might have a large, millimeter scale, 
extra dimension. In this case one might produce higher 
dimensional black 
holes in high energy particle collisions, and these might have 
experimentally accessible Hawking decay signatures \cite{willy, scott}.   

In this note we will investigate a particular set of solutions, called 
black Saturns.  However we hope that our philosophy and results 
will have some generic relevance to multihorizon black 
solutions.  A black Saturn is an exact solution of the five-dimensional 
vacuum Einstein equations, and has the horizon topology of a black ring 
that encircles a black hole.  The horizon of the ring has topology 
$S^2\times S^1$ while that of the hole is $S^3$.  The ring spins in the 
$S^1$ direction, so that centrifugal force can stabilize it.  
We also allow the hole to have angular momentum with respect to the same $S^1$.  
In fact, even if we turned off the hole's Komar angular momentum, frame 
dragging from the ring would lead to a nonvanishing angular velocity for 
the hole.  For simplicity we set angular momenta along the orthogonal, commuting 
$S^1$ to zero for both black objects.

Besides the generic motivations mentioned in the first paragraph, one
of the fundamental reasons for our interest in black Saturns is the fact that 
they offer an excellent opportunity to study the physics and thermodynamics 
of black holes with multiple horizons in asymptotically flat space. The interpretation
of multiple horizons poses interesting questions which we address below.
Trying to understand how black Saturns fit into our overall picture of black holes 
holds the potential of a deeper understanding of how thermodynamics, gravity and 
quantum mechanics fit together.

In asymptotically flat space black holes are thermally 
unstable, they Hawking radiate away their energy and angular momentum.  
So the usual way in which one studies black holes is by 
imagining that they are 
in equilibrium at a 
temperature equal to the Hawking temperature of the black hole. Since the 
presence of two horizons 
leads to two temperatures, it is not immediately clear how one should 
generalize this to our problem. One possibility is to put the 
Saturn in a reflecting box so that the ADM mass and angular momentum are 
exactly conserved, and then look for equilibria of this closed system. But this 
amounts to changing the asymptotics and might be more analogous to black 
holes in Anti de Sitter. A black Saturn may exist in 
asymptotically AdS space\footnote{But see our concluding 
discussion on what we can learn about AdS black Saturns from dual plasma 
rings and holes.}. But an exact black Saturn (or even black ring) 
solution is not yet known in AdS. Besides, when we put in a reflecting 
wall, we will have to worry about super-radiant instabilities\footnote{We 
thank Roberto Emparan for pointing this out to us.}. 
             
For these reasons, we will take another approach. 
We will consider the asymptotically flat solution as an open system. 
The question of the thermodynamics of a spacetime with multiple horizons 
is best thought of as that of a system with more than one characteristic 
temperature \cite{Maeda, Medved, paddy}, and this 
is what we will do. This means that we will be studying phases that are in {\em local} thermodynamic equilibrium. To understand the sense in which we use the word ``local", we need to look at the various timescales in the black saturn system. 
Even in associating a uniform 
temperature to a macroscopic horizon, the assumption of local equilibrium within each body is 
implicit\footnote{There are two separate 
issues here. One is that of the temperature itself, and the other 
is that of the uniformity of the temperature over the horizon. To 
identify horizon temperature with the geometrically defined surface 
gravity, we need the fact 
that it shows up in place of temperature in Hawking's Planckian 
distribution. What we are emphasizing here instead is the uniformity of 
this quantity over the horizon, to make sense of which, we 
need to distinguish competing scales.}. But there is another timescale 
that becomes relevant if the hole and the ring are to be in equilibrium 
with each other, as envisaged in \cite{feb}, while being an open system. 
For equilibrium to hold, the 
hole-ring transport has to be faster than the Hawking radiation loss. So 
for the black Saturn system we can distinguish three progressively slower 
scales - dynamics between the microscopic ingredients within a horizon, 
dynamics between the two horizons, and the Hawking leakage to infinity.




Although our black Saturns are open 
systems, we assume that the time scale associated with the energy and angular momentum loss 
to infinity due to Hawking radiation is much larger than the characteristic time 
scale of the transport which brings the hole and the ring into thermodynamic equilibrium.  Such an assumption is necessary if one wishes to study the thermodynamic phases of a composite black object, and so was implicit in the program initiated in Ref.~\cite{feb}.  There are many mechanisms in various models which might validate this  assumption, from tunneling between the black objects to interactions between their fuzz in a Mathurian approach \cite{mathur, mathur2}.
With this assumption, the black Saturn has a well-defined entropy and temperature distribution, even though it is emitting Hawking radiation.

Since Hawking radiation is by assumption slower than the hole-ring transport, our system is effectively closed and we can extremize entropy (while holding total mass and angular momentum fixed) to find the equilibrium configurations.  
This is what we set out to do. Since the second law of thermodynamics for a closed system implies that the total entropy must be non-decreasing, any global maximum of the total entropy is absolutely stable and any local maximum is metastable. Our task in this paper, then, will be to find local extrema of the total entropy of the combined hole-ring system in the moduli space of black saturn solutions with fixed ADM mass and angular momentum. At time scales where the Hawking radiation becomes important, of course, this notion of stability looses its relevance. 

As an example of a system with multiple timescales,
radiation from a heated body in local equilibrium takes away signatures of the 
temperatures in  its radiation spectrum, but that hardly means that the local 
equilibrium itself is sustained by the radiation. If a system is closed and
eventually ends up in full equilibrium, then the radiation also has to be in equilibrium, 
but otherwise, 
(massless) radiation is literally the last thing that would come to equilibrium.

We can illustrate the idea of local thermodynamic equilibrium in open systems 
using the quintessential example: a melting cube of
ice in a glass of water. For this more familiar system, around the 
neighborhood of any point we can define a local temperature. This is 
because the time scale of local 
equilibration is much smaller than the time scale of heat absorption 
from the ambiance. (In the case of the Saturn of course, the system is
losing energy, and not absorbing.). For this system, there is no detailed balancing
at all scales as there would be in the case of full equilibrium:
ice is absorbing heat from the ambiance even locally. In fact, this is the reason 
why eventually it melts! But
this does not prevent us from associating thermodynamical quantities like temperature
locally,  because local equilibrium happens through 
microscopic (molecular) processes. The region over which the temperature is 
roughly constant is determined by the relative speeds with which 
the microscopic processes and the heat absorption from the ambiance proceed. 

The idea of local equilibrium seems to be the context in which to look 
at black hole thermodynamics in general, since the picture works equally well
for holes with a single horizon. 
In fact, the conventional first and second laws of black hole thermodynamics are easily interpreted in our context.
The first law of black hole mechanics is easily applied when the system is in equilibrium, presumably with a heat bath at the Hawking temperature. The fact that black holes in flat space (which certainly do not have thermal reservoirs attached to them) are expected to satisfy this law, is already a hint that local equilibrium is how one should interpret this. Now, the second law of black hole mechanics says that $\delta {\cal A}>0$. This is usually modified to $\delta {\cal A}+\delta S_{radiation} \equiv \delta S_{\rm Total}> 0$ to take care of the Hawking effect and then interpreted as valid for the hole+radiation system. Notice that $\delta S >0$ form of the second law arises from the more general $\delta Q < T \delta S$, only under the assumption that the system is closed, $\delta Q=0$. So we see that the first law and the second law are valid as they stand as thermodynamical laws, for {\em different} thermodynamical systems: one for a black hole in equilibrium with a reservoir, the other for a closed universe with the hole in it. On the other hand, from the local equilibrium perspective, the laws of black hole mechanics are valid for the {\em same} thermodynamical system.
The price to pay is that we are looking at an open system with loss in the form of Hawking radiation and the laws of black hole mechanics capture only the other degrees of freedom.


Notice also that it is precisely because of the notion of local 
equilibrium that we are allowed to ignore the thermal quantities that
are lost to radiation.  The characteristic entropy,  
charges, {\it{etc}}. of the local phases of Saturn are much larger than those in the radiation 
emitted during the time scale required for 
local equilibration, at least in regimes where we expect classical descriptions
of black holes to hold. In particular, this means that for the purposes of local 
equilibria, we are allowed to fix the total ADM mass and angular 
momentum of the black Saturn and treat the system as effectively closed.

When we hold $M$ and $J$ fixed, from the 
explicit solution of Ref.~\cite{Saturn}, this leaves a 
2-dimensional moduli space 
of solutions\footnote{See Appendix A for the relevant formulas and the 
counting of parameters.}.  We want 
to know 
which points in this space are equilibria and of those which points are stable and which are not. 
A point in the moduli space is in equilibrium if it is a critical point 
of the total entropy with respect to the parameters. The idea of local equilibria and entropy 
maximization have been used previously in the context of (AdS) black holes in, 
for example, \cite{indra}.
One may wonder whether it makes sense to sum the entropies of 
two systems at different temperatures, but it is this sum whose variation 
is constrained to be positive by the second law of thermodynamics\footnote{This is true for closed systems. In our case, the radiated entropy during the equilibration time scale is negligible compared to the black hole/ring quantities, so the system is effectively closed. Notice also that the horizon area keeps track purely of the microscopic degrees of freedom of the black hole and not the radiation, and our aim is precisely to compute the macroscopic equilibrium configurations corresponding to those.}. Besides, the microscopic entropy of a thermodynamic system, 
thought of as its information content or the number of states, is additive over sub-systems. Roughly, one would expect that if there are (respectively) $N_1$ and $N_2$ microscopic states that realize the macroscopic states of two subsystems, 
then the macrostate of the full system can be realized in $N_1 \times N_2$ ways. Now when we take the logarithms to define entropies, we see that they
add. 
In any event, an equilibrium is absolutely 
stable if it maximizes the entropy, metastable if it locally maximizes 
the entropy and unstable otherwise. 

The technical details of our extremization problem are spelled out in Appendix B.
But it turns out that we can make many consistency checks of our approach based on 
general arguments and what is already known in the literature. We explain these below.

In Ref.~\cite{feb} the authors found balanced solutions such that the ring 
and the hole have the same temperature and angular velocity. Though the intermediate
equations are very different, we are able to reproduce these in our program. On general
grounds, it is easy to see why this should be so. 
The argument is a simple application of the first law of 
thermodynamics. It turns out \cite{feb} that classically, black Saturns satisfy a 
generalized version of the usual first law:
\beq
dM=T_HdS_H+T_RdS_R+\Omega_H dJ_H+\Omega_RdJ_R. \label{primo}
\eeq
Recall 
that the conventional first law of black hole mechanics is interpreted as a
thermodynamic law by assuming that we are looking at the black
hole as a system in thermal equilibrium at a temperature 
equal to the Hawking temperature of the hole\footnote{In this paper, we have relaxed this
to allow local equilibria, so the fact that the first law still works with two different 
sets of characteristic intensive parameters fits in naturally. 
The conventional laws of black hole mechanics arise as a succinct characterization
of the thermal nature of the microscopic black degrees of freedom when we are ignoring the
radiation. 
}. 
As we have fixed the total ADM energy, the left hand side of the 
previous equation vanishes.  At a 
critical point, assuming that the system is in local equilibrium, the total entropy
\beq
S=S_H+S_R
\eeq
is extremized and so the variations of the entropies must be opposite
\beq
dS_H=-dS_R.
\eeq
One can in fact explicitly verify numerically that such variations are allowed in
the moduli space of black Saturn solutions.
We have fixed the total angular momentum so the variations of the Komar angular momenta are also opposite
\beq
dJ_H=-dJ_R.
\eeq
Combining everything (\ref{primo}) becomes
\beq
0=dS_H(T_H-T_R)+dJ_H(\Omega_H-\Omega_R) \label{eq}
\eeq
and so, considering arbitrary variations $dS_H$ and $dJ_H$, the temperatures and 
angular velocities must be equal.  We will now argue that this argument fails along 
extremal curves on which arbitrary variations of $dS_H$ and $dJ_H$ do not exist within 
the moduli space. We have also checked that it is precisely these extra points which
we find through our approach.

\begin{figure} 
\begin{center}
\includegraphics[width=15cm]{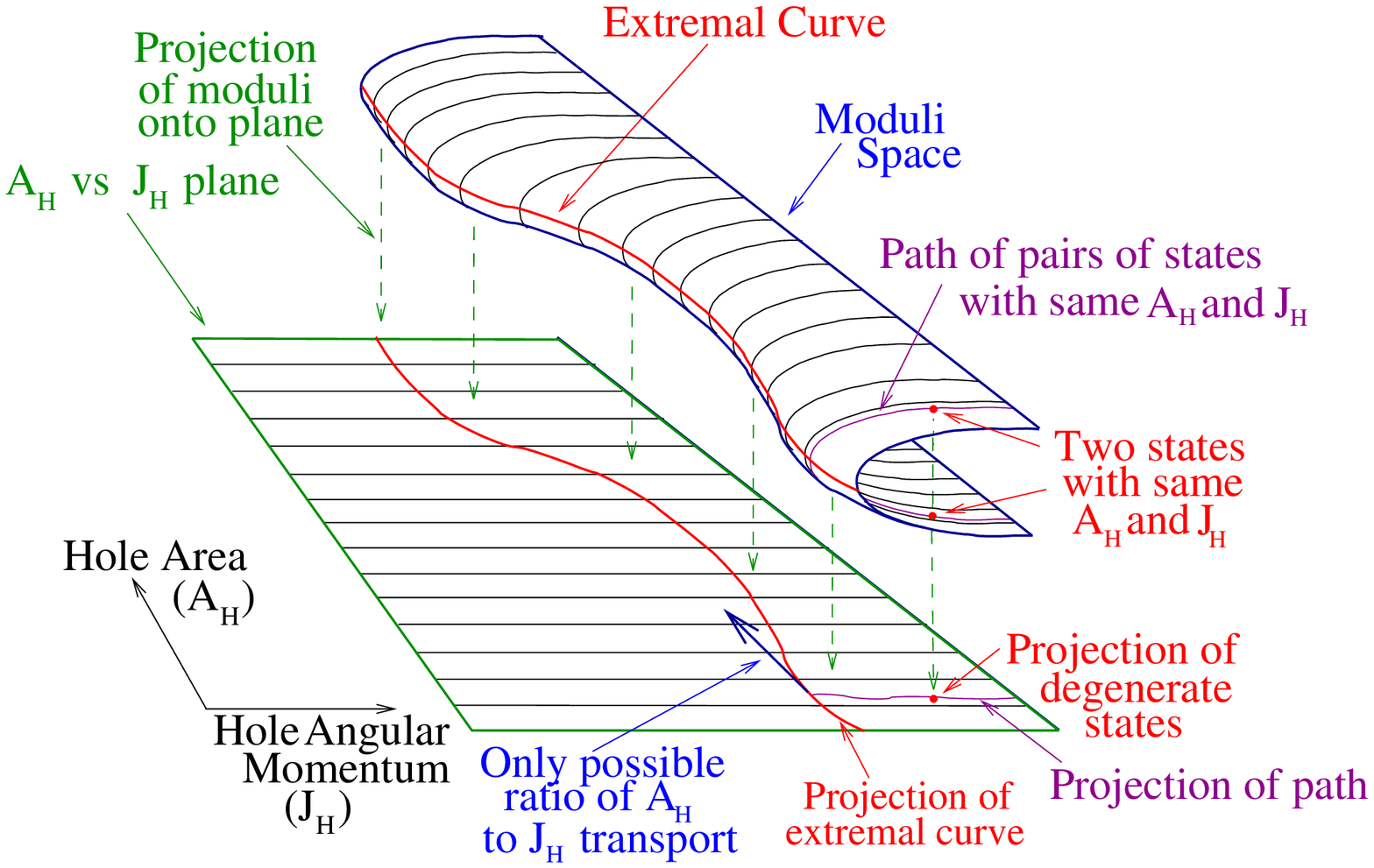} 
\end{center}
\caption{The moduli space of black Saturns that we consider is 2-dimensional.  At regular points $A_H$, the area of the event horizon of the black hole, and $J_H$, the Komar angular momentum of the black hole, are good coordinates.  However for some values of $(J_H,A_H)$ there are multiple states.  The $(J_H,\ A_H)$ coordinate system breaks down on critical curves.  When a state is on such a curve, all radiation between the hole and the ring carries a fixed angular momentum to entropy ratio from the hole.  All nonisothermal equilibria lie on such curves.} \label{crit}
\end{figure}

Locally the moduli are easy to understand, they can be parametrized by the Komar angular momentum and entropy of, for example, the hole.  We fix all other degrees of freedom by imposing rotational symmetries.  The above argument fails because globally these are not good coordinates for the moduli space.  Sometimes there are distinct configurations with the same hole angular momentum and area.  Any path between such a pair intersects a extremal curve on which these coordinates degenerate.  Consider now a path such that pairs of points reflected across the curve have the same values of angular momentum and entropy of the hole, and consider the point on which the path intersects the curve.  This situation is represented in Fig.~\ref{crit}.

On an extremal curve one cannot arbitrarily change the angular momentum and entropy of the hole, instead the entropy is minimized for each angular momentum.  To see this, notice that infinitesimal deformations tangent to the curve change the entropy and angular momentum of the hole with a fixed ratio $r$.  On the other hand, infinitesimal deformations away from the curve along the path instead leave both quantities fixed, since these quantities are the same on mirror points on the path.  Thus on the extremal curve any small deformation in this moduli space, whether or not it is tangent to the curve, changes the entropy and angular momentum of the hole with the fixed ratio $r$.  When $r$ is equal to the ratio of the temperature differences to the angular velocity differences
\beq
r=\frac{dS_H}{dJ_H}=\frac{\Omega_R-\Omega_H}{T_H-T_R} \label{conspire}
\eeq
then (\ref{eq}) is satisfied.  As Eq.~(\ref{eq}) requires $dS_H=-dS_R$, the configuration is in equilibrium even if the temperature and angular momenta differ.  We will refer to such equilibria as nonisothermal, and equilibria in which the temperatures and angular momenta are equal as isothermal.  Critical curves are codimension one in the moduli space, and the condition (\ref{conspire}) places a single additional constraint.  Thus nonisothermal equilibria occur at codimension 2, and so at isolated points in the 2-dimensional moduli space of solutions with a fixed total angular momentum and energy.

We find numerically that nonisothermal equilibria are present when the 
dimensionless angular momentum is in the window
\beq
0.914 < j < 1
\eeq
where $j$ is the reduced (dimensionless) angular momentum defined in Appendix A. 
As is implied by the above 
argument, all nonisothermal equilibria satisfy Eq.~(\ref{conspire}).  Our 
numerical results agree with this general consistency requirement.  

We will elaborate on some of the features of these non-isothermal extrema in the 
following, but before we do we emphasize some caveats.
\begin{itemize}
\item  The extrema are {\em very} non-uniformly distributed in the moduli space. In
particular, there is a lot of structure near the boundary of the moduli space. This means 
that numerically scanning for them could conceivably lose some critical points. 
We are confident 
about the curves we have found, but we do not rule out the possibility that we might 
have missed some potential critical curves.
\item  Except for the small window that we mention later, none of the critical curves are 
stable. The plots that we show below should be seen with this in mind. Some of these unstable points correspond to
local minima (both eigenvalues of the Hessian are positive) and others correspond to 
saddle points (one eigenvalue is negative, the other is positive). Along a critical curve, 
the character of the critical point can change.
\item   In 
principle there may also be critical points on the extremal curve which 
extremize both the $J_H$ and $S_H$ simultaneously.  At such points 
radiation between the hole and the ring would, to leading order, 
transport neither entropy nor angular momentum and so again the 
configuration would be an equilibrium.
\end{itemize}

One may also search for extremal curves in the phase diagram of the Myers-Perry (MP) 
black hole.  If one fixes the angular momentum of an MP black hole then one can 
always change the mass such that it is critically rotating and the area is zero.  
Thus there is no critical point where the derivative of the area vanishes.  The 
presence of the ring changes this, we find extremal curves on which the area of 
the hole's horizon is minimized and nonzero. 

\begin{figure} 
\includegraphics{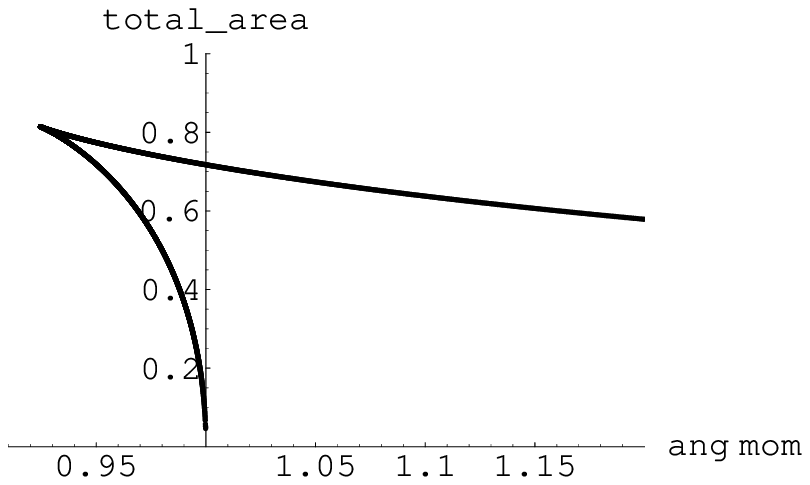} 
\includegraphics{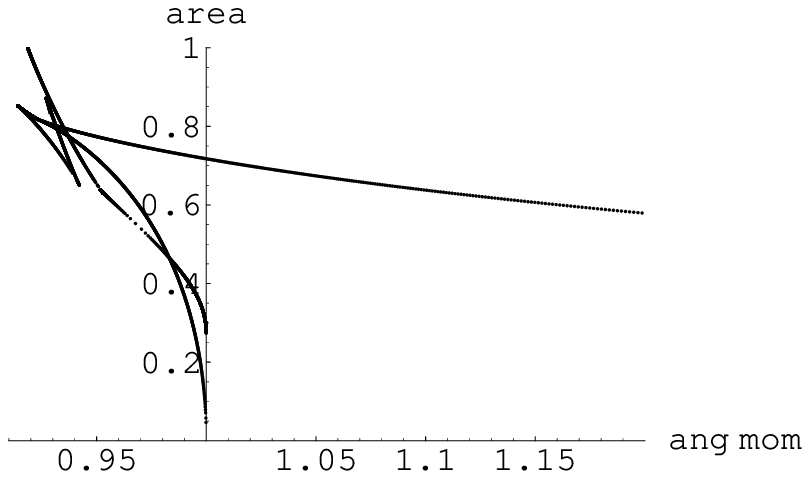} 
\caption{The moduli space of equilibria is dominated by the isothermal equilibria, shown on the left.  On the right one sees all of the equilibria, the nonisothermals are localized at small angular momentum $j$.} \label{equilib}
\end{figure}

The spectrum of isothermal critical points of black Saturn solutions was found in 
Ref.~\cite{feb}. It looks like a slightly shifted copy of the black ring phase diagram, 
with a somewhat lower entropy and slightly higher minimum angular momentum.  In 
particular, the area is maximized at the minimal angular momentum, and two branches 
emerge from this minimum, both extending to zero entropy.  One asymptotes the singular 
fat black ring at $j=1$, and the other asymptotes to the thin ring at 
$j\rightarrow\infty$.  The isothermal critical points dominate the phase diagram of 
thermal equilibria, as seen in Fig.~\ref{equilib}.  However both the equilibrium 
with the highest entropy (which is just the minimal $j$ black ring) and that with 
the lowest angular momentum are nonisothermal.

\begin{figure} 
\begin{center}
\includegraphics{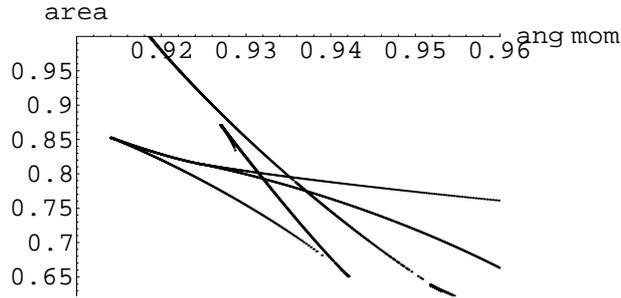} 
\end{center}
\caption{The phase diagram of equilibria (most of which are unstable) is much richer than previously thought.  Here are the lowest angular momentum equilibria, where the structure is the most intricate.  The maximum entropy is attained on the upper-left where it asymptotes to the minimally rotating black ring.  On this branch the hole is counter-rotating.  It continues to the lower-right, where the hole becomes so counter-rotating that its area eventually goes to zero at $j=1$.  The two isothermal branches go off to the right slightly higher.  They meet at the triple point $j=.92457$ and form a nonisothermal branch whose continuation yields the rest of the equilibria.  It ends when it becomes degenerate as its second derivatives vanish.} \label{noniso}
\end{figure}

The nonisothermal equilibria are concentrated around the topleft corner.  Zooming in 
on this region one finds Fig.~\ref{noniso}.  This phase diagram is surprisingly rich. 
The maximal entropy is 
considerably higher than the isothermal maximal entropy.  In fact, it asymptotes to 
the minimal angular momentum black ring at $j=\sqrt{27/32}$.  On this branch the hole 
counter-rotates, with an ever increasing counterangular momentum as the total angular 
momentum increases.  After two quick switchbacks around $j=0.952$, it continues all of 
the way to $j=1$.  The counter-rotation increases to about $-0.17$ before returning to 
$-0.16$ when $j=1$.  In this limit the hole rotates so quickly that its area and 
temperature go to zero, similarly to the extremal MP black hole.  So at the end of 
the extremal curve, the minimal area is zero, but this only occurs on the 
counter-rotating branch.  The mass of the extremal black hole is about 12 percent of 
the total mass.

The minimal angular momentum isothermal equilibrium is by no means the minimal 
angular momentum configuration.  Instead it is a triple point, where the two isothermal 
branches merge into a single nonisothermal, which continues down to $j\sim 0.914$ as 
the hole's angular momentum, mass and area increase while its temperature decreases.  
Meanwhile the ring shrinks and heats.  The endpoint appears to be the real minimum 
angular momentum for an equilibrium black Saturn.  One other branch exits from it, 
along which the hole's angular momentum and mass continue to increase.  At their 
maximum the hole accounts for only 30 percent of the mass and 20 percent of the 
angular momentum of the Saturn.  All of the equilibria are dominated by the ring.  
Anyway, this branch continues up to $j\sim 0.942$, then down to $j\sim 0.927$ and back 
to $j\sim 0.928$ as the hole shrinks again, although at the end its area grows and 
temperature decreases slightly.  The branch finally ends when the critical point 
becomes degenerate, as the second derivatives of the entropy vanish.

\begin{figure} 
\begin{center}
\includegraphics{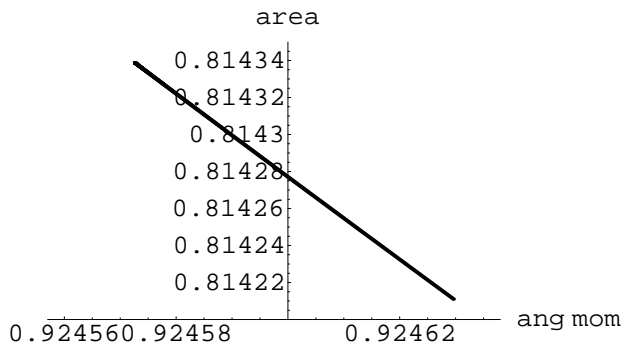} 
\end{center}
\caption{The metastable equilibria are all on the thin ring isothermal branch, in a narrow window that ends at the minimal angular momentum isothermal equilibrium, a triple point in the phase diagram where one may expect a second order phase transition.} \label{meta}
\end{figure}

None of the equilibria are stable.  This is a consequence of the fact that, as was argued in Ref.~\cite{feb}, the absolute maximum entropy occurs when all of the angular momentum is in a thin massless ring and all of the mass is in a nonspinning Schwarzschild black hole.  However, surprisingly, when $j$ is in the narrow window
\beq
0.92457<j<0.92463.
\eeq
the isothermal equilibria on the thin ring branch are local maxima of the entropy and so are metastable, at least under deformations in our moduli space.  The metastability ends precisely at the triple point, where the second derivative of the entropy functional has a zero eigenvalue, as may be expected at a second order phase transition.  The moduli space of metastable equilibria is displayed in Fig.~\ref{meta}.  

Black rings are known to suffer from radial instabilities along the fat branch and 
Gregory-Laflamme instabilities at high values of $j$ along the thin branch \cite{ringstab}. 
 It is not known whether this unstable region continues all of the way down the thin 
branch to the minimal value $j=\sqrt{27/32}$.  As has been argued in Ref.~\cite{feb}, 
one expects a similar pattern for the classical instability of the black ring in a 
black Saturn.  If black Saturns satisfy the Gubser-Mitra conjecture~\cite{indra, indra2}, 
then one expects thermodynamic stability to occur in the same range as classical 
stability.  If that is the case, our results suggest that black Saturns are stable under 
sufficiently small nonuniform radial perturbations at $j<0.92463$ on the thin ring 
branch.  This makes it more plausible that black rings themselves also have a small 
window of classical metastability at the low angular momentum end of the thin ring 
branch. Of course, Gubser-Mitra is generally known to hold only for cases with translational 
invariance, so these comments should be taken as speculative.

One observation we can make here in the Gubser-Mitra context is that if one adds an additional translationally invariant direction along which the entire solution extends trivially, then there will be more possible decay modes, as there may be decays in which there are nonvanishing derivatives along the new direction.  However the original decay modes are still present in the translationally-invariant 6-dimensional solution, they are the translationally-invariant decay modes.  Therefore the addition of a translationally-invariant direction may only increase the instability.  This implies that if a 6-dimensional translationally-invariant configuration is stable, then so is its 5-dimensional reduction.  Similarly if the 6-d solution is metastable, then the 5-d solution is either stable or metastable.  Our 5-d solution is clearly not absolutely stable, as it has a smaller entropy then a Tangherlini black hole with a thin ring, and so a metastable 6-d solution implies a metastable 5-d solution.  


All of this leads one to the question of whether or not to trust our numerical results.  Our equilibria were found independently by both collaborators, using different algorithms.  All points found not to extremize the entropy to within the precision of mathematica, about $10^{-16}$, were then thrown away.  The two authors also used different approaches to determine the metastability, one using each of the two approaches mentioned in the second appendix.  Therefore the existence and possible metastability of the points found has been well-established.  The weakness of our numerical approach is that our scans for critical points may well not have been exhaustive, and so we may well have missed some branches of critical points.

In this letter we have seen that the moduli space of black Saturn equilibria does 
not quite fit the expectations of Ref.~\cite{feb}.  In particular most of
the isothermal phases are in fact unstable. There are also extremal 
curves on which, for example, for a fixed value of the angular momenta of each black 
object, one object has an extremal area.  And on these extremal curves there are 
unstable equilibria in which the temperatures and angular velocities of the black objects are 
not equal.  We expect both of these features to be generic for composite black objects.  
In an $N$ object system with $k$ conserved quantum numbers whose totals are held fixed, 
it is always tempting to characterize the $k(N-1)$-dimensional moduli space by the area 
and $k-1$ quantum numbers of $N-1$ objects.  While this parametrization works in general 
locally, there will in general be codimension 1 extremal surfaces in the moduli space 
where the determinant of the Jacobian of the transformation from good moduli space 
coordinates to these vanishes.  

On such surfaces the ratio of the charges and entropy transported between the objects will be subject to a single constraint.  Such a constraint means that the temperature differences and potential differences may conspire, as in Eq.~(\ref{conspire}) to assure that such transport never changes the area.  All $kN-k-1$ ratios of differences of potentials need to conspire, yielding $kN-k-1$ constraints on the $(kN-k-1)$-dimensional extremal surface.  Therefore such conspiracies occur generically at isolated points in the moduli space.  In such cases the extremal configurations will be nonisothermal equilibria.  More generally there will be extremal surfaces of higher codimension, embedded in the lower-codimensional extremal surfaces, at which the possible fluxes are even more restricted.  However the same dimension-counting argument always ensures that equilibria generically occur at isolated points.

One further question we could ask is about the possibility of
black Saturns in AdS. In recent papers, Minwalla and collaborators \cite{shiraz,
shiraz2} have found 2-dimensional
dual plasma configurations (thought of as Scherk-Schwarz reduced from 4D)
corresponding to the holes and rings in AdS. Apart from the cutoff on the
angular momentum arising from the AdS radius, their phase diagrams were 
tantalizingly similar to those of the flat space black hole and black ring
shown in \cite{feb}. 
They use a rigid rotation ansatz to construct the plasma configurations, where
the configuration has uniform angular velocity. At first this seems to admit an immediate
generalization to include black Saturns, by admitting the plasma ball and the plasma ring to rotate 
independently and superimposing such configurations. 
We have done this\footnote{We do not present the computations here, because they only require minor adaptations from those of \cite{shiraz}.}, but we find that
demanding that the hole and the ring are at the same temperature and angular velocity 
eliminates the solution. So at least the plasma configurations corresponding to the
isothermal phases are not allowed in AdS.

How should we read this in the context of our previous results? We saw in the flat case that except for an extremely tiny window, all the isothermal phases are ruled out. So our results are mostly in agreement with the general features seen on the plasma side, and in particular show that the expectations from the isothermal curves of \cite{feb} should be reconsidered.
We mention here that to begin with, it is probably not legitimate to expect  more than a rough comparison between AdS and flat holes using the plasma picture: black holes in flat space are closest to small black holes in AdS
because they are unaware of the curvature of AdS. But the Minwalla et al. plasma approach is designed to handle large AdS black holes. Besides, there is the fact that AdS black holes have super-radiance, they have a new scale, their asymptotics is different, not to mention the fact that we don't have exact solutions in AdS. But despite all this, we find it encouraging that the 
expectations from plasma Saturns have some qualitative similarities with our flat space results. In particular, the window of metastability is at the very tip of the cusp, which is what would correspond to black holes at a scale much smaller than the AdS radius - precisely the place where the plasma predictions for large black holes can potentially break down.

One obvious direction here is to try a
local equilibrium analysis directly on the plasma side to see if there are phases other 
than the ones we found in flat space. The thermodynamical quantities associated with
the hole and the ring can both be written down easily for plasma following
\cite{shiraz}, so this should be a numerically tractable 
problem. 


\section* {Acknowledgement}
We would like to thank Roberto Emparan, Stanislav Kuperstein, Carlo Maccaferri and Bogdan Teaca for useful discussions. 
This work is supported in part by IISN - Belgium
(convention 4.4505.86), by the Belgian National
Lottery, by the European Commission FP6 RTN programme MRTN-CT-2004-005104
in which the authors are associated with V. U. Brussel, and by the Belgian
Federal Science Policy Office through the Interuniversity Attraction Pole
P5/27.

\appendix
\section{Black Saturn formulas}

In this section we collect the formulas associated to the black Saturn 
solution that we use for our numerics. These results are taken from 
\cite{Saturn}. The moduli space of 
black Saturns can be parameterized by three dimensionless numbers 
$\kappa_1, \kappa_2, \kappa_3$ and an overall length scale $L$. The 
$\kappa_i$ satisfy $0\le \ka_3 \le \ka_2 < \ka_1 \le 1$. The degenerate 
cases correspond to MP black holes and rings, which we are not interested 
in.
In what follows,
\bea
  \nonumber
  \bc_2 \equiv \frac{1}{\ka_2}
  \left[
   \frac{\ka_1-\ka_2}
    { \sqrt{\ka_1 (1-\ka_2)(1-\ka_3)(\ka_1-\ka_3)} } - 1
  \right] \,. 
\eea
\underline{Angular Velocities of the Horizons:}
\bea
 \Omega_H
  &=& 
  \frac{1}{L} \, \big[ 1+\ka_2\,\bc_2 \big]\,
  \sqrt{\frac{\ka_2 \ka_3}{2 \ka_1}} \, 
    \frac{\ka_3 (1-\ka_1) - \ka_1 (1-\ka_2) (1-\ka_3) \bc_2}
     {\ka_3(1-\ka_1) + \ka_1 \ka_2 (1-\ka_2) (1-\ka_3) \bc_2^2}  \, , 
\\[2mm]
   \Omega_R
  &=& 
  \frac{1}{L} \, \big[ 1+\ka_2\,\bc_2 \big]\,
  \sqrt{\frac{\ka_1 \ka_3}{2 \ka_2}} \, 
    \frac{\ka_3  - \ka_2 (1-\ka_3) \bc_2}
     {\ka_3 -  \ka_3 (\ka_1 -\ka_2) \bc_2 + \ka_1 \ka_2 (1-\ka_3) \bc_2^2}  
\, .
\eea 
\underline{Horizon Areas:}
\bea
\label{BHarea}
\mathcal{A}_H
  &=& 4 L^3 \pi^2 
  \sqrt{\frac{2(1-\ka_1)^3}{(1-\ka_2)(1-\ka_3)}}~
  \frac{1 + 
      \frac{\ka_1 \ka_2 (1-\ka_2) (1-\ka_3)}{\ka_3(1-\ka_1)} \,\bc_2^2}
             { \big( 1+\ka_2\, \bc_2 \big)^{2}}
         \, , \\[2mm]
\label{BRarea}
\mathcal{A}_R
  &=& 4 L^3 \pi^2  
  \sqrt{\frac{2 \ka_2 (\ka_2-\ka_3)^3}{\ka_1 (\ka_1-\ka_3)(1-\ka_3)}}~
  \frac{1- (\ka_1-\ka_2) \bc_2 
   + \frac{\ka_1 \ka_2 (1-\ka_3)}{\ka_3}\,  \bc_2^2  }
       { \big( 1+\ka_2\, \bc_2 \big)^{2}} \, .
\eea
\underline{Horizon Temperatures:}

\bea
T_H&=&
\frac{1}{2 L\, \pi} 
\sqrt{\frac{(1-\ka_2)(1-\ka_3)}{2(1-\ka_1)}}\,
  \frac{ \big( 1+\ka_2\, \bc_2 \big)^{2}}
     {1 + \frac{\ka_1 \ka_2 (1-\ka_2) (1-\ka_3)}{\ka_3(1-\ka_1)} 
\,\bc_2^2} \, ,   
        \\[4mm] \nonumber
T_R&=&
\frac{1}{2 L \, \pi} 
\sqrt{\frac{\ka_1(1-\ka_3)(\ka_1-\ka_3)}{2\ka_2(\ka_2-\ka_3)}}\,
    \frac{\big( 1+\ka_2\, \bc_2 \big)^{2}}
      {1- (\ka_1-\ka_2) \bc_2 
   + \frac{\ka_1 \ka_2 (1-\ka_3)}{\ka_3}\,  \bc_2^2} \, .
\eea
\underline{Komar Masses of Hole and Ring:} 
\bea
M_H &=&
  \frac{3 \pi L^2}{4 G} \,
  \frac{\ka_3 (1-\ka_1)+ \ka_1 \ka_2 (1-\ka_2)(1-\ka_3)
    \, \bc_2^2}{\ka_3(1+\bc_2\, \ka_2)} \, ,\label{mKbh} \\[2mm]  
 M_R &=&
  \frac{3 \pi L^2}{4 G} \,
  \frac{\ka_2 \big[ 1- (1-\ka_2)\, \bc_2\big]
   \big[  \ka_3 - \ka_3 (\ka_1- \ka_2)\,  \bc_2 
      + \ka_1 \ka_2 (1-\ka_3)\,  \bc_2^2\big]}
       {\ka_3(1+\bc_2\,  \ka_2)^2} \, .  \label{mKbr}  
\eea
\underline{Komar Angular Momenta:}
\bea
\label{jKbh}
  J_H &=&
  - \frac{\pi L^3}{G} \,
  \sqrt{\frac{\ka_1 \ka_2}{2 \ka_3}}\, 
  \frac{\bc_2 \big[ \ka_3 (1-\ka_1)+ \ka_1 \ka_2 (1-\ka_2)(1-\ka_3)
    \, \bc_2^2 \big]}
       {\ka_3(1+\bc_2\,  \ka_2)^2} \, , \\[2mm]
\label{jKbr}
  J_R &=&
  \frac{\pi L^3}{G} \,
  \sqrt{\frac{\ka_2}{2 \ka_1 \ka_3}}\, \\ \nonumber
  && \times
  \frac{\big[ \ka_3 - \ka_2(\ka_1-\ka_3)\, \bc_2 
              + \ka_1 \ka_2 (1-\ka_2)\, \bc_2^2 \big]
        \big[ \ka_3 - \ka_3(\ka_1-\ka_2)\, \bc_2 
              + \ka_1 \ka_2 (1-\ka_3)\, \bc_2^2 \big]}
       {\ka_3(1+\bc_2\,  \ka_2)^3} \, .
\eea

The total (ADM) mass and angular momentum are the sum of the two pieces:
\begin{eqnarray}
\label{ADMmass}
M=\frac{3\pi\, L^2}{4G}\,
  \frac{\ka_3(1-\ka_1+\ka_2)-2\ka_2\ka_3(\ka_1-\ka_2)\bc_2 
        +\ka_2\big[\ka_1-\ka_2\ka_3(1+\ka_1-\ka_2)\big]\bc_2^2}
       {\ka_3 \big[ 1+ \ka_2 \bc_2 \big]^2} \, 
\end{eqnarray}  
and
\begin{eqnarray}
\label{Jadm}
J&=& \frac{\pi\,L^3}{G}\frac{1}{\ka_3 \big[ 1+ \ka_2 \bc_2 \big]^3}
        \sqrt{\frac{\ka_2}{2\ka_1\ka_3}}        
        \, \bigg[
        \ka_3^2
        -\bar{c}_2\ka_3\Big[
(\ka_1-\ka_2)(1-\ka_1+\ka_3)+\ka_2(1-\ka_3)
                                \Big] \nonumber\\[2mm]
        &\;&\hspace{4.5cm}+\bar{c}_2^2\ka_2\ka_3\Big[
           (\ka_1-\ka_2)(\ka_1-\ka_3)+\ka_1(1+\ka_1-\ka_2-\ka_3) \Big] 
         \nonumber\\
        &\;&\hspace{4.5cm}-\bar{c}_2^3\ka_1\ka_2\Big[           
                \ka_1-\ka_2\ka_3(2+\ka_1-\ka_2-\ka_3)\Big] 
        \bigg]\, .
\end{eqnarray}

The problem we undertake is a constrained extremization of the total 
surface area ${\cal A}={\cal A}_H+{\cal A}_R$, under the condition that 
$M$ and $J$ are fixed. The computational difficulty of the problem can be 
reduced by noting that the overall scale $L$ can be eliminated by fixing the mass. Indeed, we can define
\bea
  \label{redpar}
  \begin{aligned}
  &j^2 =\frac{27 \pi}{32 G}\frac{J^2}{M^3}\;,&
  \hspace{1cm}
  &a_i =\frac{3}{16}\sqrt{\frac{3}{\pi}}\frac{{\cal
    A}_i}{(G M)^{3/2}}\;, \\[2mm]
  &\omega_i = \sqrt{\frac{8}{3\pi}}~
               \Omega_{i}(GM)^{1/2}\;,&
  \hspace{1cm}
  &\,\tau_i\, = \sqrt{\frac{32 \pi}{3}}~T_{i}(G M)^{1/2}\;, 
  \end{aligned}
\eea
and consider the equivalent problem of extremizing $a=a_H+a_R$ with fixed 
$j^2$. This means that locally there are two moduli left unfixed.
In the above, $i=H,R$. We will refer to these new variables as the reduced variables: 
their advantage is that now they depend only on the dimensionless 
parameters $\ka_i$ an not on $L$. The scaling freedom of classical general 
relativity has been used to eliminate $L$ (or equivalently $M$).

\section{Lagrange multipliers and the Hessian}

We will use Lagrange multipliers to explain our extremization. One can also work more directly by looking at the variations and constraining them, but this gives identical results (as it should).

Our aim is to extremize 
$a=a_H+a_R$ while fixing $j$ to some specific value $j_0$. So we define 
\bea
F (\ka_1,\ka_2,\ka_3, \lambda) = a(\ka_1,\ka_2,\ka_3) - \lambda \left( 
j(\ka_1,\ka_2,\ka_3)-j_0\right)
\eea
and the equations that need to be solved 
are the partial derivatives of $F$ (with respect to $\lambda, \ka_i$) set 
to 
zero:
\bea
\label{max}
a_1-\lambda j_1=0, \ \ a_2-\lambda j_2=0, \ \ a_3-\lambda j_3=0, \ {\rm 
and} \ j=j_0. 
\eea
Here, subscripts denotes partial derivatives with respect to the 
corresponding $\ka$. We can eliminate $\lambda$, and obtain a set of expressions which are immediately amenable to numerics:
\bea
\label{ext}
a_1j_2-a_2j_1=0, \ \ a_2j_3-a_3j_2=0, \ \ j-j_0=0.
\eea
To be consistent, we also need to check that the points we find this way also satisfy $a_3j_1-a_1j_3=0$. We have solved these equations using two different numerical scanning strategies and found agreement. 

The isothermal curves of \cite{feb} are obtained by 
solving
\bea
\tau_H=\tau_R, \ \ \omega_H=\omega_R, \ \ j-j_0=0.
\eea
Despite the different form that the equations take, we are also able to reproduce these curves from (\ref{ext}).

The next step is to check whether the entropy is a local maximum on the constraint 
surface. For this we have to define a notion of negative definiteness for the Hessian of $F$. If we denote the constraint 
surface\footnote{i.e.,
the space $(j-j_0)^{-1}\{0\}$, the preimage of the real number zero when $j-j_0$ is thought of as a map from the moduli space to $\IR$.} by $M$, then a bit of thought (or a computation involving changes of charts) shows that at a critical point $p$ in the moduli space which is a solution of (\ref{max}), we should study the Hessian
\bea
H=\frac{\partial^2 F}{\partial \ka_i \partial \ka_j}\Big{|}_p,
\eea
but restricted to $T_pM$. Notice in particular that there is no derivative 
with respect to the Lagrange multiplier in the Hessian. To test its negativity, we need to show that $V^T H V$ is negative for any vector $V$ lying on $T_pM$. In practice, since $H$ is a $3 \times 3$ matrix and $T_pM$ is 2-dimensional, it is more convenient to first project $H$ to a $2 \times 2$ matrix using a matrix $X$:
\bea
h=X^T H X. 
\eea
Here $X$ is a $3 \times 2$ matrix whose columns form a basis for $T_pM$.
We can construct a basis for $T_pM$ just knowing that $j-j_0$ is constant over $M$ and hence
\bea
j_1\delta \ka_1+j_2 \delta \ka_2 + j_3 \delta \ka_3=0,
\eea
where $\delta \ka_i$ can be thought of as motions along the tangent space.
From this we can extract two independent commuting vector fields, which we take to be
\bea
v_1=\left(
\begin{array}{c}
1 \nonumber \\
0 \nonumber \\
-\frac{j_1}{j_3} 
\end{array}
\right), \ \
v_2=\left(
\begin{array}{c}
0 \nonumber \\
1 \nonumber \\
-\frac{j_2}{j_3} 
\end{array}
\right).
\eea
So finally to check for local metastable equilibria, we need to check whether the matrix
\bea
h=\left(
\begin{array}{c}
v_1^T \nonumber \\
v_2^T \nonumber \\
\end{array}
\right)
\left(
\begin{array}{ccc}
F_{11} & F_{12} & F_{13} \nonumber \\
F_{21} & F_{22} & F_{23} \nonumber \\
F_{31} & F_{32} & F_{33} 
\end{array}
\right)
\left(
\begin{array}{cc}
v_1 & v_2 \nonumber \\
\end{array}
\right) \hspace{3cm} \\
=\left(
\begin{array}{cc}
F_{11}-2F_{13} \frac{j_1}{j_3}+F_{33}\big(\frac{j_1}{j_3}\big)^2 
& F_{12}-F_{13}\frac{j_2}{j_3}-F_{23}\frac{j_1}{j_3}+F_{33}\frac{j_1j_2}{j_3^2}
\\
F_{21}-F_{13}\frac{j_2}{j_3}-F_{23}\frac{j_1}{j_3}+F_{33}\frac{j_1j_2}{j_3^2} & F_{22}-2F_{23}\frac{j_2}{j_3}+F_{33}\big(\frac{j_2}{j_3}\big)^2 \\ 
\end{array}
\right) \label{mat}
\eea
has only negative eigenvalues at the critical point in question. Notice that when writing the final expression it is understood that we have solved for $\lambda$ in terms of the $j_i$ using (\ref{max}) to express $F_{ij}$
without invoking the Lagrange multiplier. This is done {\em after} everything else, because we don't want the derivatives in $\kappa_i$ to hit the $\lambda$.
While doing numerics, we should check that the values of $\lambda$ obtained from the three equations in (\ref{max}) coincide. Otherwise, it is a spurious critical point. This check automatically ensures that we are not making
any errors by counting points in $M$ where derivative(s) of $j$ with respect to $\ka_i$ vanish. We also mention that once we have (\ref{mat}) at hand
and once we have the critical points, checking the character of a critical point is numerically trivial.



\end{document}